\documentstyle[12pt,amsbsy,epsfig]{article}
\begin{document}

\date{}
\title{\normalsize{\bf{RADIATION OF PHOTONS IN PROCESS OF CHARGE PARTICLE  VOLUME
REFLECTION  IN BENT SINGLE CRYSTAL}}}

\author{ Yu. A. Chesnokov, V.I. Kotov, V.A. Maisheev and I.A. Yazynin \\ 
  \it{ Institute for High Energy Physics, 142281, Protvino, Russia }}
\maketitle
\begin{abstract}
New type of radiation in crystals is predicted and investigated in computer simulation. It is shown that process 
of volume reflection of electrons and positrons in bent crystals is accomplished with high-power 
radiation of photons. Volume reflection radiation  has intensity comparable with known 
channeling radiation, but it is less sensitive to entrance angle and sign of  charge of a particle. 
Simulated spectra of radiation power are presented for 10 GeV and 200GeV particles.
\end{abstract}
\section{Introduction}
Volume reflection of charged particles in bent single crystals was predicted 
in Ref. \cite{TV}  as result of Monte Carlo simulations. Recently this process
was observed experimentally in extracted proton beams \cite{VR}.
At volume reflection the proton crosses a number of crystallographic planes.
Due to this fact its motion has a complicated oscillating character.
For this reason  one can expect valuable radiation losses of energy for light particles 
(such as electrons and positrons).

In this paper we investigate the radiation of energetic  photons by electrons 
(positrons) moving in bent single crystals at conditions of volume reflection effect. 
Our consideration based on the equations derived from quasi-classical operator
method (see Ref.\cite{BKS}). Within this method the probabilities of QED processes
may be expressed by way of classical trajectories of charged particles in 
electrical fields.

Taking into account the above mentioned we start from consideration of
motion of charged particles in  bent single crystals and then we calculate the
differential energy losses of radiation process. 
It should be noted that the analytical description of volume reflection
one can find in paper \cite{VM}.

\section{ Motion in bent single crystals}

One can
describe the motion of  ultrarelativistic particles in  bent single crystals
with the help of the following equations\cite{VM} :
\begin{equation}
  E_0 \beta^2 v_r^2/(2c^2) +U(r)+ E_0 \beta^2 (R-r)/r= E=const,
\end{equation}
\begin{equation}
dy/dt=v_y=const,
\end{equation}
\begin{equation}
v_z = rd\phi/dt \approx  c(1-{1\over 2\gamma^2}-{(v^2_r + v_y^2)\over 2c^2}).
\end{equation}
These equations take place for the cylindrical coordinate system ($r, \phi, y$). 
Here $v_r$ is the component of particle velocity in the radial direction, $v_y$ is the component
of the velocity along $y$-axis and $v_z$ is the tangential component of the velocity,
$R$ is the radius of bending of a single crystal, $E_0$  and $\gamma$ are the particle energy and
its Lorentz factor, $E$ is the constant value of the radial energy, $U(r)$ is the one dimensional potential
of a single crystal, $c$ is the velocity of light and $\beta$ is the ratio of the particle velocity
to  velocity of light. In this paper we consider the planar case when the scattering 
is due to the interaction of particles with the set of the crystallographic planes located normally to ($r, \phi$)-plane. 
On the practice it means that $v_y/c \gg \theta_{ac}$ but $v_y/c \ll 1$ for ultrarelativistic particles,
where $\theta_{ac}$ is the critical angle of  axial channeling.

Eq.(1) one can transform in the following
form:
\begin{equation}
E_0 \beta^2 v_x^2/(2c^2) + U(x) +E_0\beta^2 x/R=E,
\end{equation}
where $x$ is the local Cartesian coordinate  which connected with the cylindrical coordinate $r$
through the relation  $x= R-r$ and $v_x=v_r$.  We also changed $r$-value in the denominator of Eq.(1) on $R$.
For real experimental situation it brings a negligible mistake (of the order of $ x/R$).
In Eq.(4)  $E$-value have a sense of the transversal energy.

The considered here equations describe three dimensional motion of  particles in a
bent single crystal in the cylindrical coordinate system.  Let us introduce the Cartesian coordinate
system in which the $xy$-plane is coincident with the front edge of  a single crystal. Now
we can calculate  $x$-component of the velocity in this system:
\begin{equation}
V_x= v_x \cos\phi +v_z\sin \phi \approx  v_x+v_z \phi.
\end{equation}
\begin{equation}
V_z= -v_x \sin\phi +v_z\cos \phi.
\end{equation}
Fig.1 illustrates the geometry  of particle  volume reflection and different  coordinate systems
which will be used in our consideration.

With the help of Eq. (5) we can calculate the velocity $V_x$ as a function of time.
For specific calculations we select the energy of electron (positron) beam equal to 200
GeV and (110) plane of the silicon single crystal. The planar potential
for this plane calculated on the basis of results of x-ray diffraction (see details in Ref.{\cite{VM}}). 
Fig. 2 illustrates the behavior of velocity $V_x$ at symmetrical orientation of
single crystal (when entering angle of particle is equal to exit angle).
The radius of the curvature is equal to 10 meters and the thickness of the single crystal
is approximately equal to 0.06 cm.
One can see that particle performs an aperiodic oscillations in the transversal plane.

\section{Radiation energy losses of particle}

For calculations of radiation energy losses of particle we use the following relation
(see \cite{BKS}):
\begin{equation}
{d{\cal{E}} \over dE_\gamma}= {i \alpha m^2 c^4 \over 2\pi \varepsilon^2} \omega
\int_{\bf{D}} {dt d\tau \over \tau-0}\{1+{{\varepsilon^2 +{\varepsilon^{'}}^2}\over {4 c^2\varepsilon \varepsilon^{'}}}
\gamma^2  [\Delta {\bf{v}}(t- \tau/2)-\Delta{\bf{v}}(t+\tau/2)] ^2\} \exp{-iA_1}\,  ,
\end{equation}   
\begin{equation}
A_1={{\omega \varepsilon \tau }\over {2 \varepsilon^{'}}}
[{1\over \gamma^2}+{1 \over \tau} \int _{-\tau/2}^{\tau/2} ds (\Delta{\bf{v}}(t+s)/c)^2
-({1\over \tau} \int_{-\tau/2}^{\tau/2} ds \Delta {\bf{v}} (t+s)/c)^2],
\end{equation}  
where $\Delta {\bf{v}}(t, {\bf{v}}_0) ={\bf{v}}(t_1) -{\bf{v}}_0  $ is the
 velocity variation as a function of time $t_1$, $m$ and $\gamma$ are the mass and Lorentz factor of particle,
 $E_\gamma, \omega$ are the energy and frequency of photon, $\varepsilon $ is the particle energy, $\varepsilon^{'} =\varepsilon- E_\gamma$. The time variables $t_1$ and $t_2$
($t_2$ is  time variable as  $t_1$) connected with variables $t$ and $\tau$
by equations: $t_1=t-\tau/2$ and $t_2=t+\tau/2$. ${\bf{D}}$ is the domain of 
definition of integrand function.  
Eq. (7) has a common character and may be used for  a solution of a numerous
radiation problem. In particular with the help of this equation relations were obtained which
describe the radiation at quasiperiodic motion, synchrotron like radiation
and so on \cite{BKS}. For these aims some simplifications of the integrand functions were made.
Most difficult case for calculations is the case when the angle of deflection of
moving  particle on the formation length  is compared with the characteristic radiation
angle  ($1/\gamma$).

Consideration of the curves in Fig. 2 shows  that the character of motion at volume reflection 
significantly differs from motion in a straight single crystal. 
The peculiarities are: aperiodicity of oscillations and deflection of particles (electrons)
of order of characteristic radiation angle. The time for one oscillation changed
from $\approx 50$ fs (femptosecond) near critical point till $ \approx 20$ fs
on the entrance (exit) of single crystal. The value of characteristic radiation angle
(for condition in Fig.2) is equal to $\approx 2.5 \mu rad$. The variation of relative transversal
particle velocity (corresponding changing of angle) is $\approx 10^{-5} - 2 10^{-6}$.
Let us recall the following relation :
\begin{equation}
\omega={2\gamma^2 \omega_0 \over 1+\rho/2},
\end{equation}
(where $\rho=2\gamma^2 \overline{v^2_\bot}$  and $v_\bot$ is the transversal particle velocity)
which defined a frequency of emitted photon by the frequency $\omega_0$ of particle motion.
In our case we should take the time $T_0$ of every oscillation and calculate
frequency $\omega_0 =2\pi /T_0$. Taking into account that parameter $\rho \sim 1$
we get estimation for photon energy $ \sim \gamma^2\omega_0$. 
Thus we see that more intensive range of photon spectrum (for condition as in Fig.2)
corresponds to $15 -35$ GeV.  
Another obvious peculiarity of the radiation at volume reflection is dependence 
the form of photon spectrum  on the thickness of bent single crystal.
Really with increasing of the thickness the number of periods also increases
and their period is decreased. In this case one can expect increasing of number
of high energy photons. 

The calculation of radiation losses spectra with accordance Eq.(7) is serious
computional problem, mainly due to oscillating character of the integrand function.
However, we do not see clear possibility for  a simplification of Eq.(7). We think
that in this case the direct calculations with the use of Eq. (7)  may be very useful for further
understanding of process and allow one to get necessary simplifications.

For calculation of differential radiation energy  losses we create program,
which allow us to find photon distributions in accordance with Eq. (7).
We made computations for radiation of 200 GeV electrons and positrons
in thick silicon bent single crystal for conditions shown in Fig. 2.
It should be noted that calculations are very  time consuming. This time depends also
on the accuracy of calculations. Our experience shows that for the same accuracy
the calculations at higher energy require more time.

One can expect that the integration over $\tau$ in Eq. (7) may be performed in
limits of time which necessary for formation of radiation. Partially, it is so.
Really, main contribution in losses come from this range. However,
there are  oscillations of integrand function at large enough values $\tau$ which
can change result on 5-15 \%. If do not take this into account the integral 
may be calculated fast but accuracy will not high. Nevertheless, it may be
possible way for simplification of Eq. (7).

Fig. 3 illustrates the result of calculation of differential radiation energy losses
for 200 GeV electrons.  According to our estimation the accuracy of this
calculation is 1-2\% for photon energies less then 45 GeV, and  about 5\%
from 45 GeV till 85. 
As expected the main losses located in the range 15-40 GeV.

The differential energy losses of positrons presented in Fig. 3 show that the form of photon spectrum
is similar (in comparison with radiation of electrons) but intensity in range  less than 45 Gev
for positrons is approximately larger on $\approx 20 \%$. We estimate accuracy
of calculations for this spectrum as 10-15 \%.
 Fig. 4 illustrates the differential radiation energy losses for 10 GeV positrons
which move in the (110) plane of 0.045 cm silicon single crystal with the 5 m bending radius.

 Note that the photon spectrum should be averaged over
transversal energies as it was described in \cite{VM}. However, for large enough radii
of bending (in comparison with the characteristic radius \cite{VM}) the averaged spectrum 
is approximately the same as calculated for
one transversal energy. 

In calculations we consider the pure process in a planar electric field of a single crystal
and hence we do not take into account the thermal fluctuations of the potential
and multiple scattering by atoms. Thermal fluctuations are the cause of
appearance of amorphous contribution in the photon spectrum (see, for example \cite{BKS, BM}).
One can expect that the intensity of this process is close to similar one
in straight single crystals. It means that in the range 20-40 GeV
radiation at volume reflection exceed the amorphous contribution more
than 30-40 times.

The multiple scattering can change the probability of radiation if angle of the 
scattering is close or exceed the characteristic angle of radiation on the
formation length. It is easy to estimate that for conditions in Fig. 2
the angle due to multiple scattering (on the formation length)
 is less than 1$\mu$rad for photons with energies more than 10 GeV and 
therefore multiple scattering does not influent significantly on the radiation process. 

It should be noted that Eq. (7)  does not take into account multiple processes
of photon radiation. However, presented in this paper photon spectra
may be used for  calculation of these processes. 

\section{Conclusion}

    It should emphasized in conclusion   volume reflection radiation  is  in tens times more  intensive than  radiation in  
amorphous material. This new kind of radiation favorably differs also from radiation at channeling \cite{ME}, 
because there is  no sharp dependence on beam angular  spread  and  sign of  charge of a particle.
For these reasons volume reflection radiation has exclusive perspectives for innovation in electron accelerators for creation 
of powerful radiation sources. Another possible application of volume reflection radiation is high-energy hadron and 
electron identification in physic experiments due to strong dependence of it parameters from particle 
mass. 

The described type of radiation can be observed in radiation experiments, 
which are started at IHEP U-70\cite{VTB} and CERN SPS \cite{VR,JU}.

  We are grateful to Profs.  N.E.Tyurin and A.M.Zaitsev for the support and fruitful 
discussions. We acknowledge partial support by the Russian Foundation for Basic Research grants 
05-02-17622, 07-02-00022 and INTAS-CERN Foundation (grant No.05-103-7525).

\newpage
\begin{figure} 
\begin{center}
\parbox[c]{14.5cm}{\epsfig{file=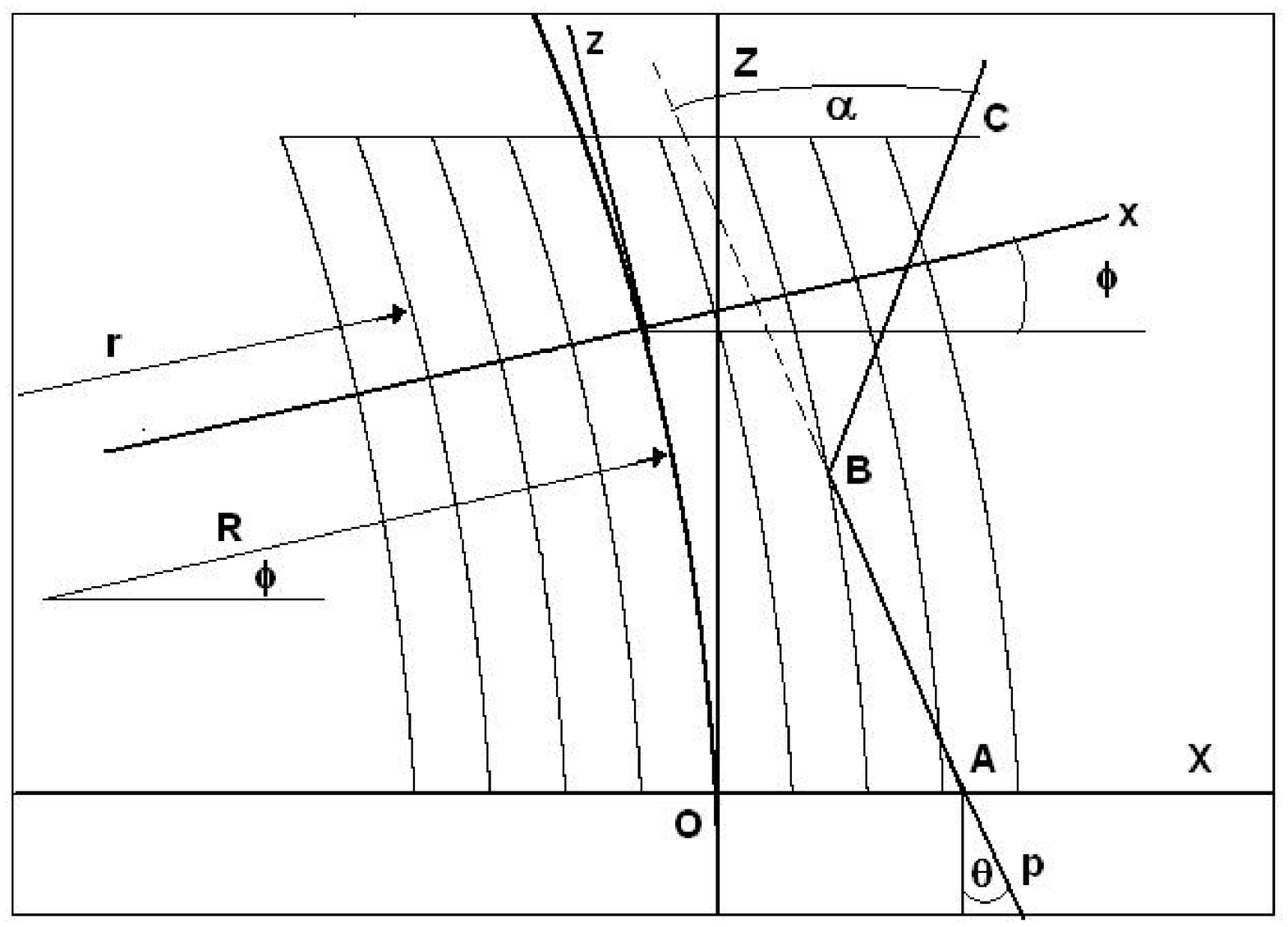,width=12cm}}
\parbox[c]{15cm}{\caption{Scheme of the volume reflection of the proton beam. $XYZ$ is the Cartesian coordinate system at the
entrance in single crystal, $xyz$ is the local Cartesian coordinate system connected with the current location 
of the particle. $Y$-axis is directed normally to the plane of figure. $\theta$ and $\alpha$ are the initial and  volume
reflection angles.   
              }}
\end{center}
\end{figure}
\newpage
\begin{figure} 
\begin{center}
\parbox[c]{14.5cm}{\epsfig{file=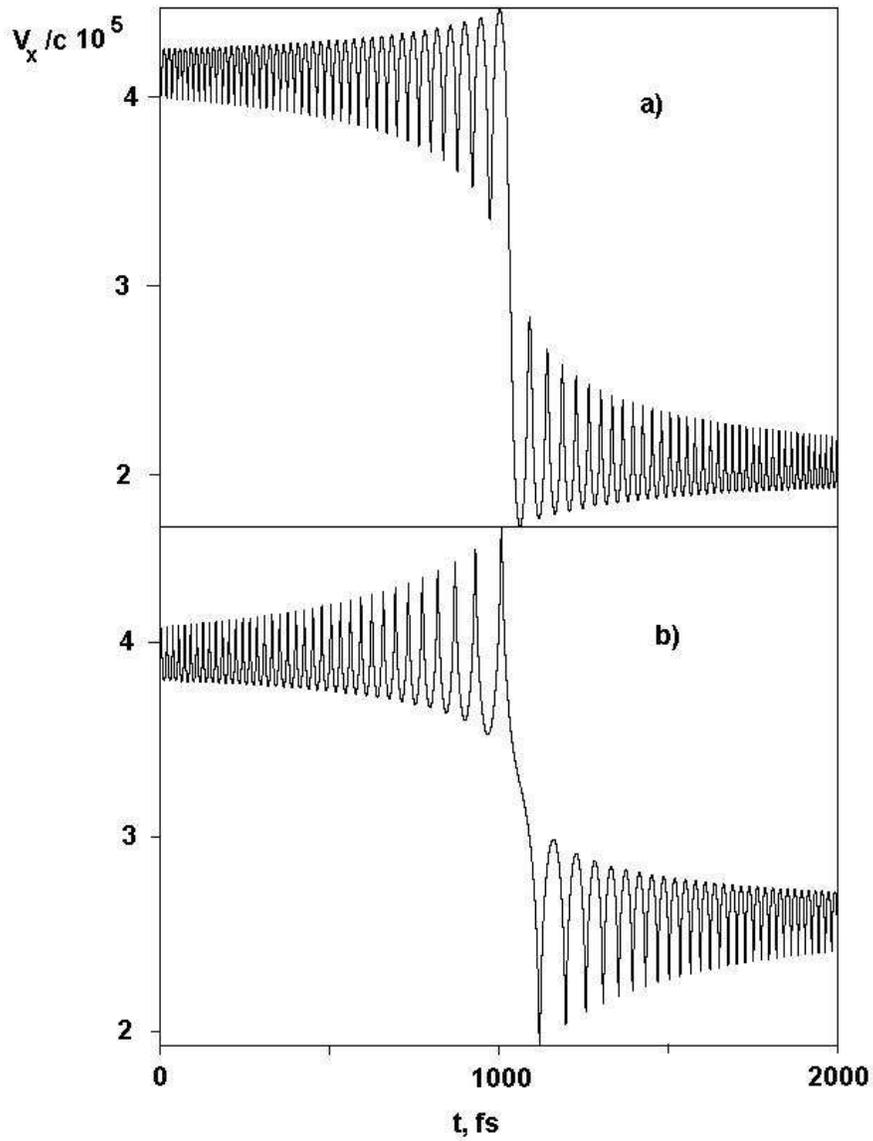,width=12cm}}
\parbox[c]{15cm}{\caption{Relative transversal velocities of pozitrons (a) and
electrons (b)  at volume reflection in (110) plane of bent silicon single crystal
as a functions of time.
Particle energy is equal to 200 GeV, crystal thickness is 0.06 cm, radius of bending is 10 m.
              }}
\end{center}
\end{figure}
\newpage
\begin{figure} 
\begin{center}
\parbox[c]{14.5cm}{\epsfig{file=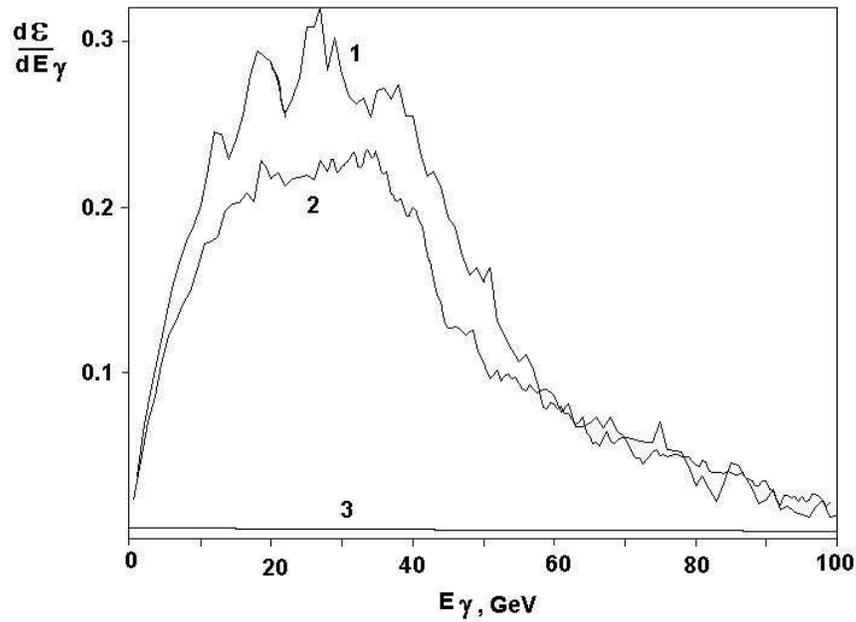,width=12cm}}
\parbox[c]{15cm}{\caption{Differential radiation energy losses 200-GeV positrons (1) and
electrons (2) for conditions as in Fig. 2. The curve 3 is amorphous contribution.
              }}
\end{center}
\end{figure}
\newpage
\begin{figure} 
\begin{center}
\parbox[c]{14.5cm}{\epsfig{file=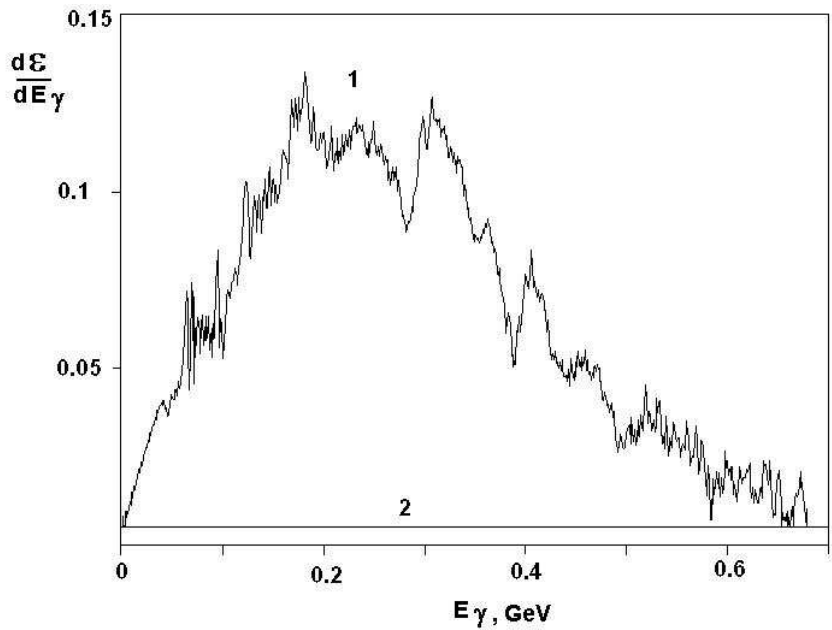,width=12cm}}
\parbox[c]{15cm}{\caption{Differential radiation energy losses 10-GeV positrons (1)
  in 0.045 cm silicon single crystal. The curve 2 is amorphous contribution.           }}
\end{center}
\end{figure}
\end{document}